\documentclass[aip,jcp,amsmath,amssymb,preprint]{revtex4-1}

\usepackage{graphicx}
\usepackage{dcolumn}
\usepackage{bm}
\relax

\begin{document}
\title{A Moving-Trap Zeeman Decelerator}
 \author{Lewis A. McArd}
 \affiliation{Joint Quantum Centre Durham-Newcastle, Department of Physics, Durham University, South Road, Durham DH1 3LE, United Kingdom}
 
 \author{Arin Mizouri}
 \affiliation{Joint Quantum Centre Durham-Newcastle, Department of Physics, Durham University, South Road, Durham DH1 3LE, United Kingdom}
 
 \author{Paul A. Walker}
 \affiliation{Joint Quantum Centre Durham-Newcastle, Department of Physics, Durham University, South Road, Durham DH1 3LE, United Kingdom}
 
 \author{Vijay Singh}
 \affiliation{Joint Quantum Centre Durham-Newcastle, Department of Physics, Durham University, South Road, Durham DH1 3LE, United Kingdom}
 
 \author{Ulrich Krohn}
 \affiliation{Joint Quantum Centre Durham-Newcastle, Department of Physics, Durham University, South Road, Durham DH1 3LE, United Kingdom}
 
 \author{Ed A. Hinds}
 \affiliation{Centre for Cold Matter, Blackett Laboratory, Imperial College London, Prince Consort Road, London SW7 2AZ, United Kingdom}
 
 \author{David Carty}
 \email{david.carty@durham.ac.uk}
 \affiliation{Joint Quantum Centre Durham-Newcastle, Department of Physics, Durham University, South Road, Durham DH1 3LE, United Kingdom}
 \affiliation{Joint Quantum Centre Durham-Newcastle, Department of Chemistry, Durham University, South Road, Durham DH1 3LE, United Kingdom}

 \date{\today}

\begin{abstract}

We present a moving-trap Zeeman decelerator (MTZD) for use in molecular beam manipulation and magnetic-trapping experiments of paramagnetic atoms and molecules. The 0.49~m MTZD consists of a combination of a 2D magnetic quadrupole guide, for transverse confinement, and deceleration coils for longitudinal confinement, which together produce an array of 3D magnetic traps. The design is based on that of Trimeche \emph{et al.} with significant modifications. The 2D quadrupole is driven by fast rising and falling square pulses of current (up to 700~A) of arbitrary lengths of time. The traps can be made to move at velocities from \emph{ca.}~370~m~s$^{-1}$ down to zero, which is done by driving through the deceleration coils a sinusoidal current with frequencies ranging from zero to \emph{ca.}~9~kHz and peak currents up to 1000~A. The trap velocity is proportional to the current frequency. The MTZD manipulates the velocities of molecular beams by traveling initially at the same velocity as the beam. 3D guiding is achieved with a constant current frequency and deceleration is achieved with a downward chirp of the current frequency at a rate corresponding to the desired deceleration. We explain the technical design and operation principles of the MTZD and we detail the custom power electronics that drive the 2D quadrupole guide and the decelerator coils. We present extensive Monte-Carlo trajectory simulations to demonstrate the properties of the MTZD and we conclude that decelerations should be kept below $3\times10^4$~m~s$^{−2}$ to maintain a good 6D phase-space acceptance. In proof-of-principle experiments, with the deceleration coils operating with a reduced current in the range 100--200~A, we demonstrate the 3D guiding of a beam of metastable argon ($^3P_2$) atoms at 373~m~s$^{-1}$ and deceleration at $2.5\times10^4$~m~s$^{-2}$ from 342~m~s$^{-1}$ to 304~m~s$^{-1}$. The deceleration corresponds to the removal of 21\% of the kinetic energy of the beam.

\end{abstract}


 \maketitle

\section*{Introduction}

There are are many potential applications for molecules that have been cooled to cold and ultracold temperatures. These can be categorised into three ``grand challenges'': \emph{precision measurement}; \emph{quantum simulation} and \emph{controlled cold and ultracold chemistry}. Precision measurement is where molecules can be used to test fundamental physics beyond the standard model. For example, molecules can be used in the measurement of the dipole moment of the electron,\cite{Baron2014} or in measurements of the time variations of fundamental constants \cite{Truppe2013,Jansen2013,Cheng2016} or in the measurement of parity violation in chiral molecules.\cite{Tokunaga2017} Quantum simulation is where a controllable array of mutually interacting quantum entities, such as ultracold dipolar molecules, is used to simulate real, many-body quantum systems that are impossible to simulate using classical computation.\cite{Blackmore2018} Such problems are found in condensed matter physics, quantum chemistry and cosmology.\cite{Georgescu2014}

Controlled cold and ultracold chemistry is where fundamental chemical collision dynamics becomes dominated by quantum effects, such as tunnelling through potential barriers, and where the dynamics can be manipulated using external fields.\cite{Krems2008} At temperatures below a few tens of mK, the de Broglie wavelength of the reactants is comparable to the length of characteristic intermolecular interactions. This can lead to the formation of metastable collision complexes due to quantum tunneling through a centrifugal barrier, especially when the collision energy is tuned to a quasi-bound resonance state on other side of the barrier; this can drastically affect the rates of elastic, inelastic and reactive collisions. The energies of quasi-bound resonance states can be altered by the application of external electric, magnetic or electromagnetic fields, thus allowing control over collision dynamics.

An example of controlled cold and ultracold chemistry is the sympathetic cooling of molecules in conservative traps. Sympathetic cooling is where cold molecules thermalise to ultracold temperatures through elastic collisions with laser-cooled, ultracold atoms. In a conservative trap, such as an electrostatic Paul trap or a magnetic quadrupole trap, molecules are in low-field-seeking states. Such states are not the absolute ground state, which is  high-field-seeking, therefore inelastic collisions to the absolute ground state will lead to trap loss. Inelastic collisions can occur just as frequently as elastic collisions, thus making sympathetic cooling impossible. Some systems have been identified where sympathetic cooling may be theoretically possible.\cite{Wallis2009,Wallis2011,Hummon2011,GonzalezMartinez2013} One example is the sympathetic cooling of CaH with laser-cooled Li. CaH can undergo a barrierless, exothermic chemical reaction with Li,\cite{Singh2012} but in a magnetic field and with the CaH and the Li in their maximally spin-stretched states, chemical reactivity and inelasticity may be controlled so that elastic collisions dominate and the molecules are cooled to ultracold temperatures.

Recently, rapid advances have been made in the direct laser cooling and magneto-optical trapping of molecules at ultracold temperatures. Thus far, SrF,\cite{Barry2014} YO,\cite{Hummon2013} CaF,\cite{Zhelyazkova2014} YbF,\cite{Lim2018} and SrOH\cite{Kozyryev2017} have been laser cooled. SrF\cite{Steinecker2016} and CaF\cite{Williams2017,Anderegg2017} have been captured in 3D molecular magneto-optical traps (MOTs) and subsequently loaded into magnetic traps.\cite{Williams2018,McCarron2018} Anderegg \emph{et al.} captured $10^5$ CaF molecules at a density of $7\times10^6$~cm$^{-3}$ at 340~$\mu$K.\citep{Anderegg2017} Sub-Doppler cooling to 50~$\mu$K has been demonstrated for CaF producing $2\times10^4$ molecules at a density of $2.5\times10^5$~cm$^{-3}$.\cite{Truppe2017} Currently, molecular MOTs are loaded from a slow buffer-gas beam source followed by laser slowing, which is an inefficient method that appears to be limiting MOT densities. A more efficient loading method using a technique called Zeeman-Sisyphus deceleration has been proposed for CaF, which uses a combination of spatially varying magnetic fields from a \emph{ca.}~1~m array of permanent magnets and optical pumping to decelerate the CaF.\cite{Fitch2016} Zeeman deceleration, which uses time-varying magnetic fields produced by electromagnets to slow pulsed beams of paramagnetic atoms or molecules to arbitrary low velocities, would also be an efficient and general way to slow packets of molecules to below the capture velocity of a MOT, which is typically around 10~m~s$^{-1}$.

Zeeman decelerators have been developed to produce controlled atomic and molecular beams for applications in high-resolution spectroscopy, low-temperature trapping experiments and high collision-energy-resolution scattering experiments that make use of the narrow velocity spread, velocity control and state purity of the output packet. Multi-stage Zeeman decelerators, in which a series of switched solenoids provide longitudinal deceleration as well as transverse focusing, suffer from losses in particle density due to strong coupling between the longitudinal and transverse oscillatory motions that cause parametric amplification of particle trajectories. Losses are particularly punishing at velocities below around 100~m~s$^{-1}$. To mitigate the effects of losses in multi-stage Zeeman decelerators, several advanced modes of operation have been devised.\cite{Wiederkehr2010,Dulitz2014,Toscano2017} However, while these advanced strategies work well at high velocities, they have not solved the problem of large losses over the full range of velocities down to zero. In scattering experiments where a high collision-energy resolution is desirable, there is little need for velocities below around 100~m~s$^{-1}$ as the collision energy is set by the velocities of both colliding beams and their crossing angle. Low collision energies can be reached by reducing the crossing angle, although this does come at the expense of resolution in the collision-energy. However, for experiments such as sympathetic cooling or molecule-MOT loading, where achieving as high a density of molecules as possible is imperative, efficient deceleration to velocities around 10~m~s$^{-1}$ is necessary.

Traveling-wave decelerators overcome the limitations of multi-stage Zeeman decelerators by employing a 3D magnetic trap that moves with the same initial velocity as the molecular beam pulse and gradually decelerates to an arbitrary final velocity. The particles remain confined in the moving traps at all times, which is in contrast to multi-stage Zeeman decelerators which provide a time-averaged trap.

In this paper we will present the improved design of the travelling-wave Zeeman decelerator originally reported by Trimeche \textit{et al.}.\cite{Trimeche2011}  A different design has been adopted by the Narevicius group.\cite{Lavert-Ofir2011} Briefly, the decelerator is comprised of four modules of flattened helical coils and a quadrupole guide to produce a moving 3D potential. The design and the two types of power electronics required to drive the high currents through the helical coils and the quadrupole will be described and their performance analysed. This will include detailed Monte-Carlo trajectory simulations of the dynamics of the trap and the effects this has on the acceptance of the decelerator. The experimental results of the decelerator loaded with metastable argon will be presented. Here we will show the 3D guiding (or velocity bunching) and deceleration of metastable argon in a traveling-wave decelerator and compare this to Monte-Carlo trajectory simulations.

\section*{Principles of Zeeman decelerators}\label{sec:background}

Zeeman decelerators make use of the force experienced by a paramagnetic atom or molecule by a time-dependent magnetic field. The operation of a multistage Zeeman decelerator was originally reported independently by Vanhaecke \textit{et al.} and Narevicius \textit{et al.} in 2007.\cite{Vanhaecke2007, Narevicius2007a} The deceleration process relies on the field generated by a series of solenoids (deceleration stages) to generate a quasi-moving potential. Along the axis of the solenoid, the magnetic field is highest in the centre and paramagnetic particles that enter the decelerator have their longitudinal kinetic energy converted into Zeeman potential energy. Particles in low-field-seeking states experience an increase in potential energy as they traverse the solenoid and are thus decelerated. Before the particle reaches the centre of the solenoid (and would begin to accelerate out the other end) the field is switched off thus permanently removing the kinetic energy that has been lost. Multiple stages are used to further reduce the kinetic energy of the particle.

When describing this type of Zeeman decelerator, it is convenient to describe the longitudinal position of the particles in terms of a phase angle $\phi$ due to the periodicity of the deceleration process. A phase angle of 90$^\circ$ corresponds to the centre of the solenoid while a phase angle of 0$^\circ$ corresponds to the centre of the space between two adjacent solenoids. The amount of kinetic energy removed from a particle per coil is dependent on the particle's position along the beam axis at the point the magnetic field is switched off. This occurs once a synchronous particle, defined as a particle travelling along the beam axis with longitudinal velocity $v_0$, reaches a chosen equilibrium phase angle, $\phi_0$. By definition, the synchronous particle will have the same amount of kinetic energy removed per stage. A particle starting from the same spatial location as the synchronous particle, but with a higher longitudinal velocity, will travel further into the solenoid before the field is switched off. Over a series of deceleration stages, this asynchronous particle will gain in phase angle until it acquires a longitudinal velocity of $v_0$. At this point the relative phase angle between the asynchronous and synchronous particles begins to decrease. The asynchronous particle will return to the same equilibrium phase angle as the synchronous particle but this time with a lower longitudinal velocity. The opposite process then occurs to complete an oscillation around the synchronous particle; this is the concept of phase stability. To complicate matters, the transverse field of the solenoid varies as a function of longitudinal position. Close to the centre of the solenoid the transverse field is concave, leading to the focusing of a packet of asynchronous particles surrounding the synchronous particle. However, at large distances from the centre of the coil, the field is convex and this defocuses the packet of particles. This relationship between the longitudinal and transverse fields couples the longitudinal and transverse motion of the particles, particularly at low velocities. The effect this has on the 6D acceptance of the decelerator has been studied by Wiederkehr \textit{et al.} \cite{Wiederkehr2010} and generally results in a lowering of the acceptance of the device as many trajectories through the decelerator become phase unstable. To overcome this issue, improvements to the design and operation of the decelerator have been made. For instance Dulitz \textit{et al.} have demonstrated that periodically using two adjacent solenoids in an anti-Helmholtz configuration can provide some transverse confinement and increase the number of particles at the end of the decelerator.\cite{Dulitz2014} Similarly, the decelerator developed by Cremers \textit{et al.}  alternates the  decelerator solenoid stages with permanent hexapole magnets to improve transverse confinement.\cite{Cremers2017}

Another solution to phase instabilities has been to redesign the Zeeman decelerator so that the particles are confined longitudinally and transversely throughout the deceleration process within a moving 3D magnetic trap. This technique is analogous to traveling-wave Stark decelerators.\cite{Osterwalder2010} To date, a design of traveling-wave, or moving-trap, Zeeman decelerator has been reported by Lavert-Ofir \textit{et al.} \cite{Lavert-Ofir2011a} and by Trimeche \textit{et al.}.\cite{Trimeche2011} Although the principle of each is largely the same, the coils used to generate the fields are different as are the techniques used to generate the time-varying currents in the coils. In both designs, the molecular beam traverses a vacuum tube and the coils sit outside vacuum. The Lavert-Ofir \textit{et al.} design has much the larger 6D acceptance  because the vacuum tube has an internal diameter of 10~mm compared to 1~mm in the Trimeche \textit{et al.} design. The velocity acceptances are broadly similar.

The decelerator presented here is based on the design developed by Trimeche \textit{et al.}, but we have modified it substantially to increase the 3D spatial acceptance without compromising the 3D velocity acceptance. The advantage of the Trimeche \textit{et al.} design is that it allows, in principle, access through the coils to the vacuum tube, which is potentially beneficial for additional vacuum pumping in a long decelerator.

\begin{figure}
\centering
\includegraphics[width=360pt]{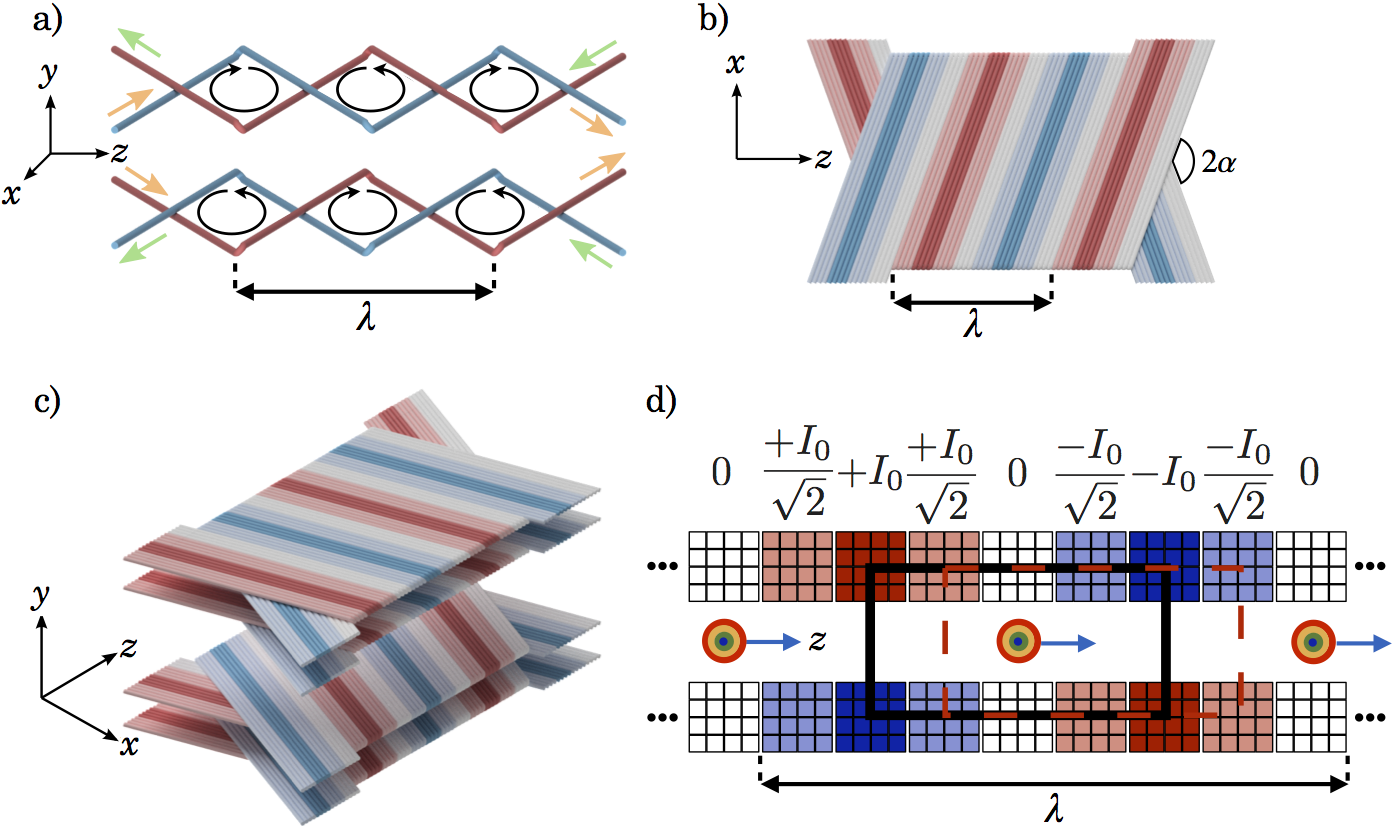}
 \caption{\textbf{The schematic representation of the construction of the planar helical coils.} Panel a) shows a simplified pair of flattened helical coils consisting of just four wires. The coils are offset in the y direction while remaining symmetrically spaced around the z axis. The lower set of coils is wound with the opposite handedness to the upper coils so as to form a periodic anti-Helmholtz array. The arrows represent the direction of the current flow in the wires. The spatial period of the coils, $\lambda$, has been labelled. Panel b) shows a two-period coil constructed from two sets of 16 wires (coloured red and blue). The half angle at which the wires cross, $\alpha$ is also shown. Panel c) shows the exploded view of a complete two period coil module. The module is formed by a total of four coils, two above the z axis and two below. Panel d) shows a cut through the $zy$-plane of the module shown in panel c). Only the first nine trap poles are illustrated. The poles are coloured to show which poles have a common phase of the current. The poles are labelled with the value of the current flowing through that pole at the point in time in the phase when the current is at the maximum value, $I_0$. These poles are linked with a solid black rectangle to highlight how those four poles are arranged in a quadrupole. The broken red line shows where the trap minimum will be at a time one quarter of a phase later. Adjacent poles are labelled with their current values relative to $I_0$. \label{fig:coil_geometry}}
\end{figure}

A simplified schematic of a coil module is shown in Figure~\ref{fig:coil_geometry}a). The simplified coil module is constructed from just four wires that form two sets of helical coils which are equidistantly spaced around the molecular beam axis along $z$. The coils below the $z$ axis are wound with the opposite handedness to those above the $z$ axis. The coils have a spatial periodicity of $\lambda$ and, as a result, the magnetic field produced is also periodic (hence the field is often referred to as a wave). The period of the coil is given by

\begin{equation}
\lambda = \frac{32d_\mathrm{w}}{\mathrm{sin\left( \alpha\right) }},
\end{equation}

\noindent where $d_\mathrm{w}$ is the diameter of the wire and $\alpha$ is the half angle the wires cross at.

The current flow through each wire element has been annotated on to Figure~\ref{fig:coil_geometry}a). Tracing the current around a single coil it is possible to see that the periodically crossing wires form a geometry which can be approximated to a set of rhombus shaped rings with a circulating current. Since the current in the lower set of rings flows in the opposite direction to that through the upper rings, an array of adjacent anti-Helmholtz coil pairs each approximating a quadrupole trap is formed. To increase the number of effective turns, each of the wires already shown is duplicated and laid next to the original but translated slightly along the $z$ axis. This is repeated another twice giving a four-fold increase in the magnetic field generated. To create a series of overlapped quadrupole traps, a duplicate of the four-turn quadrupole trap array is made and laid next to the original again translated along the $z$ axis. This is repeated a further two times until the wire arrangement, or coil, shown in Figure~\ref{fig:coil_geometry}b) is constructed. This coil is effectively a flattened double helix of two 16-wire ribbons (shown in different colours). The number of effective turns is then further doubled by stacking a duplicate of each coil onto the originals so that a pair of complete coils lies above and below the $z$ axis, as shown in Figure~\ref{fig:coil_geometry}c). Each simplified quadrupole trap in Figure~\ref{fig:coil_geometry}a) then ultimately consists of 16 turns. A cut through the $yz$ plane of the coils as viewed along the $x$ axis is shown in Figure~\ref{fig:coil_geometry}d). At a given point in time, $t_0$, blocks of 16 wires, each carrying a current of $16\times I_0$, make the poles of a quadrupole trap. Such a quadrupole trap is highlighted in Figure~\ref{fig:coil_geometry}d). Also highlighted is an adjacent, overlapping quadrupole trap whose poles carry a current at $t_0$ with a value $16\times I_0/\sqrt{2}$. The next adjacent overlapping quadrupole trap carries no current at $t_0$ and the next one carries a current $16\times I_0/\sqrt{2}$ at $t_0$. These four overlapping traps make one spatial period of the decelerator. The quadrupole traps are made to move along the molecular beam axis by varying the currents in the poles sinusoidally in time with an angular frequency $\omega$. Effectively, as one trap is ramping off the next adjacent trap is ramping on such that the magnetic field zero moves smoothly along $z$. This concept is similar to that used by Greiner \textit{et al.} used to transport samples of ultracold atomic vapours over large distances.\cite{Greiner2001} The velocity, $v_z$, at which the trap moves along the $z$ axis is proportional to $\omega$ and the coil period $\lambda$,

\begin{equation}
v_\mathrm{z} =\frac{\lambda\,\omega}{2\pi} = \frac{16\,d_\mathrm{w}\,\omega}{\pi\sin(\alpha)}.
\label{equ:vel_frequ}
\end{equation}

The coils are wound from AWG 18 ($d_\mathrm{w}=1.2$~mm), kapton-wrapped, copper wire. The coil modules are encased in thermally-conducting epoxy resin and constructed around a stainless steel tube with an internal diameter of 5.35~mm and 0.5~mm wall thickness. A half angle $\alpha=70^\circ$ was chosen giving a coil period of $\lambda=40.9$~mm. A coil module consists of three periods. When $I_0=500$~A, the depth of a trap along the $z$ axis is approximately 0.7~T.

\begin{figure}[t]
\centering
\includegraphics[width=340pt]{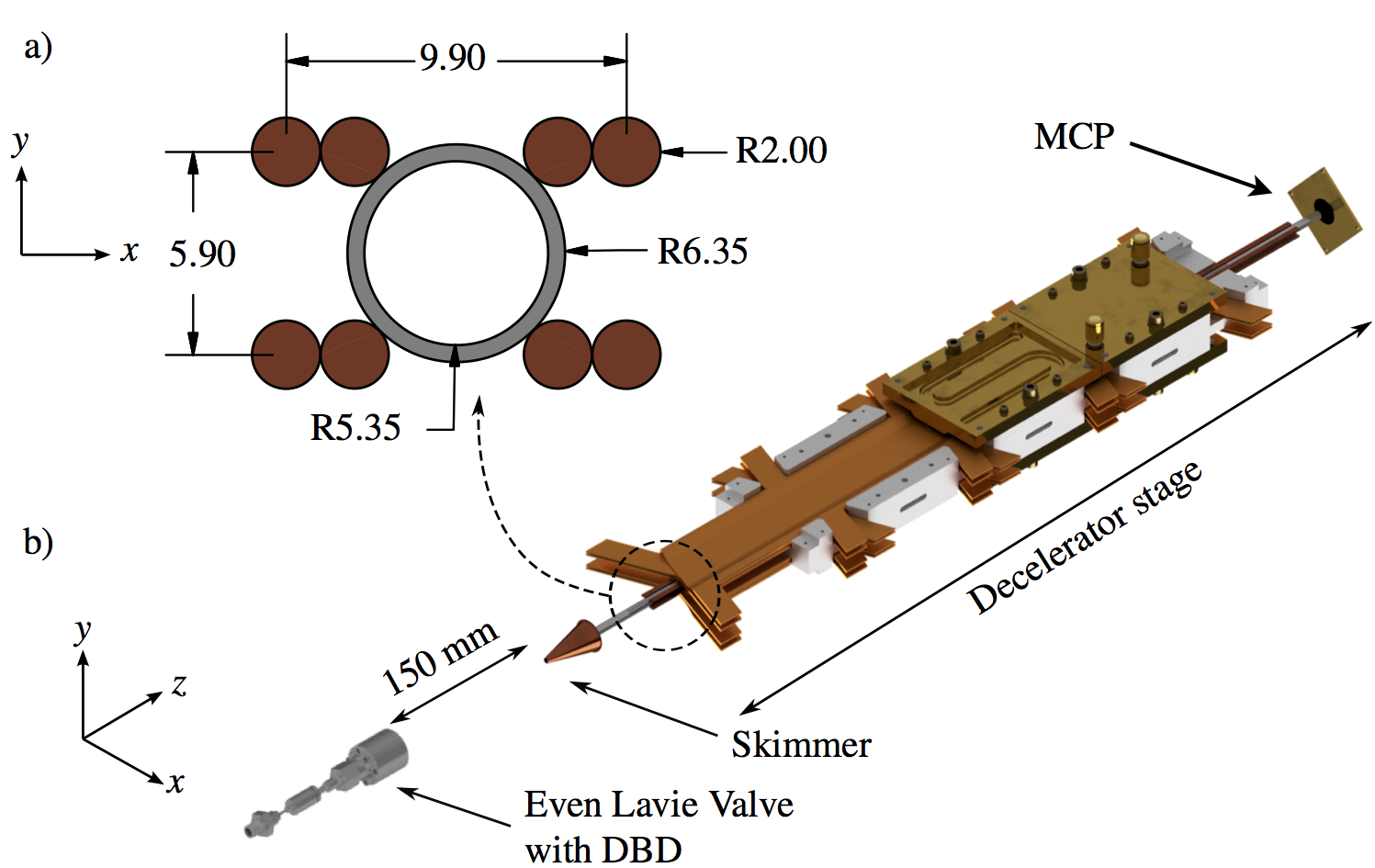}
 \caption{\textbf{A simplified diagram of the decelerator with a cross-sectional view of the quadrupole guide.} Panel a) shows the geometry of the quadrupole in the xy plane. The dimensions (in mm) of the poles and stainless steel tubing have been annotated. Panel b) shows a drawing of the parts of the decelerator including the Even-Lavie valve, the DBD, the skimmer and MCP. This drawing also details the construction process of the coil modules. Both the quadrupole and helical coils are supported by aluminium blocks and are cooled with water which flows through the brass plates above and below the helical coils.\label{fig:complete_dec}}
\end{figure}

The trap is very shallow in the $x$ direction, but non-zero due to the half angle $\alpha$. A quadrupole guide running down the entire length of the decelerator is included to increase the trap depth in the $x$ direction. Each pole of the quadrupole guide consists of two, 2 mm diameter, solid copper wires arranged as shown in cross section in Figure~ \ref{fig:complete_dec}a). With a DC current of 700~A in each wire, the trap depth along the $x$ and $y$ axes due to the combined field produced by the decelerator coils and the quadrupole guide is 0.2~T and 0.5~T, respectively. Figure~\ref{fig:complete_dec}a) shows that the quadrupole guide is centred around the stainless steel tube and is located between the upper and lower pair of deceleration coils. The geometry of the complete four-module, \emph{ca.}~0.5~m-long decelerator is shown in Figure~\ref{fig:complete_dec}b). Figure~\ref{fig:complete_dec}b) also shows that each coil module sits in an aluminium mount and that the mount supports brass plates through which chilled water flows to cool the coil and mount.

\begin{figure}
\centering
\includegraphics[width=340pt]{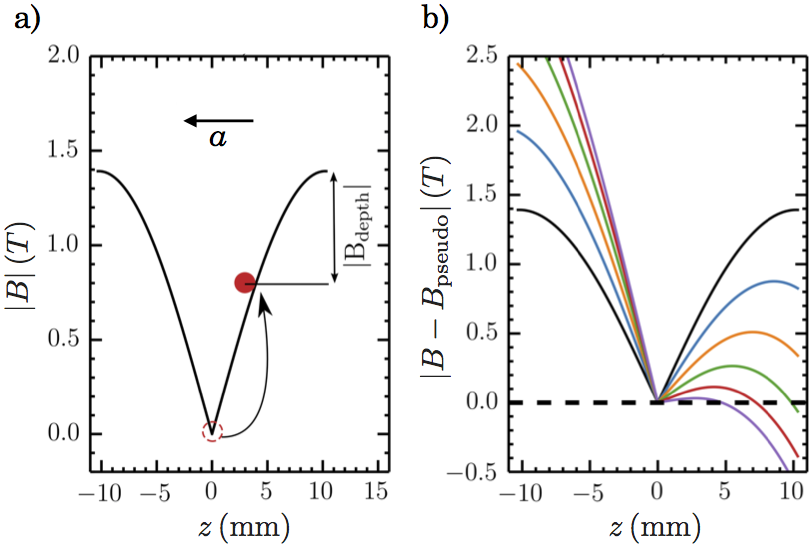}
 \caption{\textbf{Two representations of the consequence of the deceleration on the potential experienced by a particle.}
      Panel a) shows a representation of the synchronous particle of the trap. The trap as viewed in an inertial frame is shown in black. At a constant trap velocity the synchronous particle will coincide with the trap minimum (red dashed dot). The deceleration results in the synchronous particle rising up the trap edge resulting in a lower effective potential depth. Panel b) shows an alternative view expressed in terms of a pseudo force experienced by the particles. The example shows how the field in the inertial frame (black) generated at 1000 A travelling at 300 m$\,$s$^\mathrm{-1}$ is modified by the deceleration. Here, metastable argon ($^3P_2,m_J=2$) is considered. The curves show the effective field for a series of final trap velocities of 260 m$\,$s$^\mathrm{-1}$ (blue), 220 m$\,$s$^\mathrm{-1}$ (orange), 180 m$\,$s$^\mathrm{-1}$ (green), 140 m$\,$s$^\mathrm{-1}$ (red) and 100 m$\,$s$^\mathrm{-1}$ (purple). Again demonstrating the lowering of the front potential edge.}\label{fig:pseudo}
\end{figure}

Applying a constant-frequency, sinusoidal waveform of current in each adjacent overlapped trap, where the waveform in a given trap is $\pi/4$ out of phase with the waveform in the previous trap, results in a series of traps separated by $\lambda/2$ that move at a constant velocity. A constant deceleration of the traps can be achieved by linearly chirping-down the frequency of the current waveforms. A synchronous particle travelling at the same velocity as a trap will always remain in the same position in the trap as the trap moves. A non-synchronous particle will oscillate around the synchronous particle, provided they remain in the stable region of phase space. At constant trap velocity, the synchronous particle will be located at the trap centre, but during deceleration the synchronous particle experiences a force that pushes it up the potential in the forward $z$ direction closer to the trap edge, as shown in Figure~\ref{fig:pseudo}a). The force is the equivalent of a pseudo-magnetic field, $B_\mathrm{psuedo}$, given by

\begin{equation}
B_\mathrm{pseudo} = \frac{a\,m\,z}{\mu},
\end{equation}

\noindent where $a$ is the acceleration, $m$ is the particle mass, $z$ is the position of the particle measured from the trap centre and $\mu$ is the magnetic moment of the particle. The magnetic field, $|B_\mathrm{depth}|$, that must be overcome for the particle to escape from the trap is shown in Figure~\ref{fig:pseudo}a) and defines the effective depth of the trap during deceleration. In this example, the trap was calculated using a current in the deceleration coils of $I_0=1000$~A. The larger the deceleration, the shallower the trap depth and the lower the phase-space acceptance of the trap. Figure~\ref{fig:pseudo}b) shows an alternative view of the lowering of the trap depth for various values of deceleration, where the pseudo-magnetic field has been subtracted from the magnetic field of the trap to give an effective trap magnetic-field profile that is proportional to the effective potential experienced by a particle in the inertial frame of the particle. The particle being decelerated has a mass of $m=40$~u and a magnetic moment of $\mu=3~\mu_\mathrm{B}$, where $\mu_\mathrm{B}$ is the Bohr magneton. In Figure~\ref{fig:pseudo}b), the initial speed of the trap in each example was 300~m~s$^{-1}$ and the deceleration was applied over a distance of 49~cm to final velocities of 260~m~s$^{-1}$, 220~m~s$^{-1}$, 180~m~s$^{-1}$, 140~m~s$^{-1}$ and 100~m~s$^{-1}$.

\section*{Experimental apparatus}
\subsection*{Molecular beam}
A pulsed supersonic expansion provides the starting point for this experiment. Here, an Even-Lavie valve is employed, which is mounted in a copper cooling jacket. This is fed with a supply of liquid nitrogen in order to cool the valve. The valve is fitted with a dielectric barrier discharge (DBD) which can be used to produce molecular radicals. However, in this experiment it is used to produce a beam of argon atoms in the metastable $^3P_2$ state, which has long lifetime. The valve is mounted 150 mm away from the conical skimmer which has an orifice diameter of 4 mm. This has been annotated onto Figure~\ref{fig:complete_dec}b) which shows a simplified view of the decelerator. These parameters were chosen to reduce skimmer interference with the beam.\cite{Even2014} Ar($^3P_2$) has 11.54 eV of internal energy,\cite{Futch1956} which allows it to be detected directly using a Micro-Channel Plate (MCP) detector. The MCP sits some 30 mm away from the end of the quadrupole guide and 92.4 cm from the valve nozzle.

\subsection*{Quadrupole power electronics}
As shown in Figure~\ref{fig:complete_dec}a), each pole of the quadrupole guide consists of two wires, each carrying up to 700~A of DC current. In actual fact, the quadrupole is made from one single wire that is wound continuously, which means that at the beginning and end of the quadrupole guide the wire loops from one pole to the other in the $x$ or $y$ directions. This creates a fringe field at the beginning and end of the quadrupole guide that is difficult to model and has an unknown effect on the particle trajectories as the particles enter and leave the quadrupole guide. To avoid this, the quadrupole guide should be switched on only once the pulse of particles has fully entered the quadrupole guide and should be switched off before the pulse leaves the quadrupole guide.

The quadrupole-guide electronics supply switchable DC currents up to 700 A through the quadrupole guide for an arbitrary length of time. Two high-power supplies (Ametek Sorensen, SGA40-375D, 15kW), each capable of delivering 375 A at 40 V, are connected in parallel. The problem with this type of current source is that it cannot be switched on and off quickly enough to couple the particle pulse into and out of the quadrupole guide effectively. The 0--90\% rise time is 209 $\mu$s and the 100--0\% fall time is 362 $\mu$s when the supplies are connected to a 63~cm-long quadrupole guide. The switching times can be improved by incorporating a circuit inspired by that used by Wiederkehr \textit{et al.}.\cite{Wiederkehr2011} The core component of this circuit is a capacitor charged to a modest voltage. By simply discharging the capacitor through the quadrupole one can generate rapidly brief current pulses of almost 1000 A. The rate at which current is drawn is dependent on the properties of the resistance and inductance of the quadrupole guide and the capacitor. Through careful choice of circuit layout the capacitor can provide a large voltage bias to the quadrupole guide to drive the current off. Again, the new off time is dependent on the properties of the load and capacitor.

\begin{figure}[t]
\centering
\includegraphics[width=350pt]{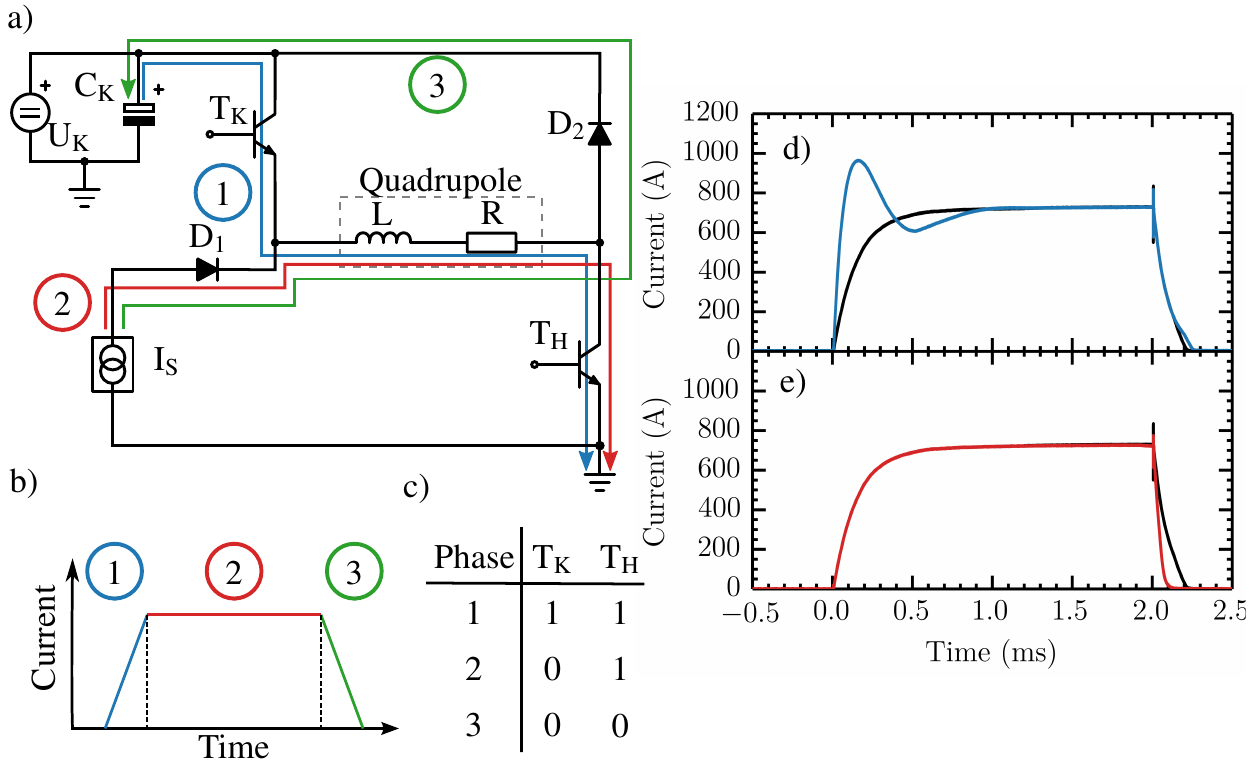}
 \caption{\textbf{The circuit diagram and principle of operation of the quadrupole guide power electronics.}
      Panel (a) shows a circuit diagram of the quadrupole guide power electronics, where the quadrupole guide coil is represented by an inductor and a resistor. Current flow through the quadrupole guide coil is broken into three phases. The flow of the of the current pulse in each phase has been sketched on the circuit. Panel (b) shows an idealised view of the current flowing through the quadrupole guide coil as a function of time and the phase it corresponds to. The Kick supply, $\mathrm{U_K}$, is used to charge the kick capacitor $\mathrm{C_K}$. The IGBTs, $\mathrm{T_K}$ and $\mathrm{T_H}$, are used to select which phase the circuit is in. A summary of the logic states the IGBTs are in is shown in the table in panel c). Here, 1 means active, thus in a closed switch position, and 0 means inactive, thus in an open switch position. The hold supply is labelled $\mathrm{I_S}$. The three phases are the kick (blue), hold (red), and decay (green). Panels (d) and (e) show current pulses of 700 A with hold times around 650 μs. The black curve shows the pulse without the kick circuitry. The blue curve shows how the rise time of the pulse is modified when only the kick and hold phases are activated. the red curve shows how the fall time is modified when only the hold and decay phases are activated.}\label{fig:quad_circuit}
\end{figure}

Figure~\ref{fig:quad_circuit}a) shows a schematic of the quadrupole-guide power electronics. The schematic has been annotated to include the direction of current flow during three phases of the pulse: the `kick' phase, which induces the rise in current to switch the quadrupole guide on; the `hold' phase, during which current is drawn from the power supplies only and the `decay' phase, which drives the current out of the circuit to switch the quadrupole guide off. The resulting current pulse is sketched in Figure~\ref{fig:quad_circuit}b).

In order to switch between the three phases of the current pulse, a pair of Insulated Gate Bipolar Transistors (IGBTs) were employed (Semikron SKM600GB066). The state of the IGBT during each phase has been summarised in the table in Figure~\ref{fig:quad_circuit}c). Not shown in the figure are the control electronics consisting of the IGBT-driver cards (Semikron SKHI 10/12 (R)) and the fibre-optic coupler to an FPGA that controls the timing. The kick phase begins with the activation of both IGBTs T$_\mathrm{K}$ and T$_\mathrm{H}$. This allows the quadrupole to draw a current from the kick capacitor which is charged by the supply U$_\mathrm{K}$ (TDK Lambda 202A-3kV-POS-PFC). The diode D$_1$ prevents current flowing to the lower voltage hold power supply. As the current extracted from the kick capacitor may exceed the current rating of the IGBT, the circuit limits the current by rapidly switching the transistor T$_\mathrm{K}$ to maintain an average current. This also limits the voltage to which the capacitor discharges to the value required during the decay phase.

During the hold phase, the flow of current is controlled by T$_\mathrm{H}$. After an arbitrary hold time after which T$_\mathrm{H}$ is deactivated, the quadrupole guide will try to maintain the field produced by the current following Faraday's law. To increase the rate of this inductive decay, the load is biased by the remaining kick capacitor voltage from earlier to drive the decay of current through diode D$_2$ back into the capacitor. Figures~\ref{fig:quad_circuit}d) and \ref{fig:quad_circuit}e) show current pulses of 700 A with hold times around 650 $\mu$s. The kick voltage was initially set to 60 V, although it can be set up to 90 V. In Figure~\ref{fig:quad_circuit}d), the black curve shows the pulse without the kick circuitry as measured using an open-Hall effect current probe (Pico Technology, TA167 AC/DC current probe). The blue curve shows how the rise time of the pulse is modified when only the kick and hold phases are activated. The current rose to a peak of 963 A before settling to the hold current of 700~A. The rise time was reduced to 58.4 $\mu$s. Figure~\ref{fig:quad_circuit}e) shows how the pulse is modified when only the hold and decay phases are activated. With the remaining voltage in the kick capacitor at 60 V, the fall time was reduced to 178 $\mu$s. This second mode of operation has been used during the acquisition of the experimental results because the rear fringe field is far more detrimental to the metastable argon signal detected on the MCP.

\subsection*{Decelerator power electronics}

The power electronics used for the decelerator coils must be able to supply pure or chirped sinusoidal waveforms with peak currents up to 500~A at frequencies varying between zero and approximately 10 kHz to each of the four independent coil phases. The impedance of the coils is such that an emf of 800~V is required giving a peak power of 283~kW. The difficulty with building power supplies capable of generating an alternating current is that the impedance of the positive and negative voltage-controlling devices would have to vary over time. Such devices would have to dissipate the power by heating so their use is restricted to low power applications. At high powers, this leaves a voltage-controlling device that has two states (an `on' state with resistance close to zero and an `off' state with very high resistance) such as an IGBT at ones disposal. These would minimise power dissipation, but would preclude their use in generating a pure sinusoidal waveform. However, the technique of pulse-width modulation (PWM) can be used to generate current waveforms that are approximately sinusoidal by taking advantage of the exponential rise and decay of currents in inductive and resistive loads, as seen with the current pulses in the quadrupole guide in Figure~\ref{fig:quad_circuit}d).

\begin{figure}[t]
\centering
\includegraphics[width=340pt]{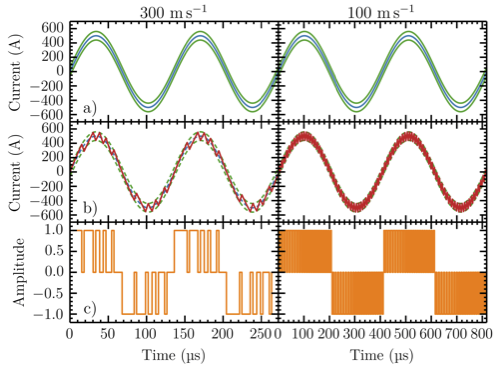}
 \caption{\textbf{The process used to synthesise a sinusoidal waveform for two example sine waves.} The example waveforms have frequencies equivalent to a trap velocity of 300 m$\,$s$^{-1}$ and 100 m$\,$s$^{-1}$ both with an amplitude of 500 A. Row a) shows the first step in generating the waveform. A reference waveform (blue) is generated from which an envelope (green) is created by offsetting the target waveform vertically by a fraction of the target peak current. Row b) shows the synthesis of the current waveform (red) which is generated through the application and removal of DC voltage pulses. Row c) shows the voltage pulses required to generate the current.\label{fig:PWM}}
\end{figure}

In PWM, a series of square voltage pulses of varying widths and constant amplitude are used to synthesise the waveform. To begin calculating the sequence of voltage-pulse widths, a ``target waveform'' is generated. A ``tolerance envelope'' is generated above and below the target waveform by offsetting the target waveform vertically upwards and downwards by a fraction of the target peak current. The combination of the target waveform and the tolerance envelope is shown in Figure~\ref{fig:PWM}a). The left and right panels show waveforms required to make moving traps of 300~m~s$^{-1}$ and 100~m~s$^{-1}$, respectively, with frequencies determined by equation~\ref{equ:vel_frequ}.

Starting from the beginning of the waveform, a voltage pulse is applied to allow the current to exponentially rise up to the point the current meets the upper limit of the tolerance envelope. The time this takes determines the width of the first voltage pulse and depends on the response characteristics of the coil. Therefore,the response characteristics must be known before the sequence can be calculated. At this point the voltage is off and the current exponentially decays until the current meets the lower limit of the tolerance envelope. The voltage is then applied again for long enough to allow the current to rise to the upper limit of the tolerance envelope. When the voltage is off, the current decays to the lower limit again. The negative cycle of the sine wave is generated in the same way, but the sign of the voltage applied across the coil is reversed so that the current flows in the opposite direction. By repeating this process the complete pulse sequence that will generate the desired waveform will be calculated. It should be noted that IGBTs tend to need a short recovery period before they can be switched again that must be considered during the pulse-sequence calculation. Figure~\ref{fig:PWM}b) shows a current waveform generated using the pulse sequence shown in panel c) and an analytical form of the measured rising and falling current response characteristics of the coils.

For a given set of response characteristics of the coil, high-frequency waveforms require fewer voltage pulses in the sequence than low-frequency waveforms. Therefore, high-frequency target waveforms are approximated less well by the synthesised waveform than low-frequency target waveforms. The highest frequency target waveform that can be synthesised by PWM is one that requires a single voltage pulse for each half cycle. However, the synthesised waveform will be a poor approximation of the target waveform.

\begin{figure}[t]
\centering
\includegraphics[width=340pt]{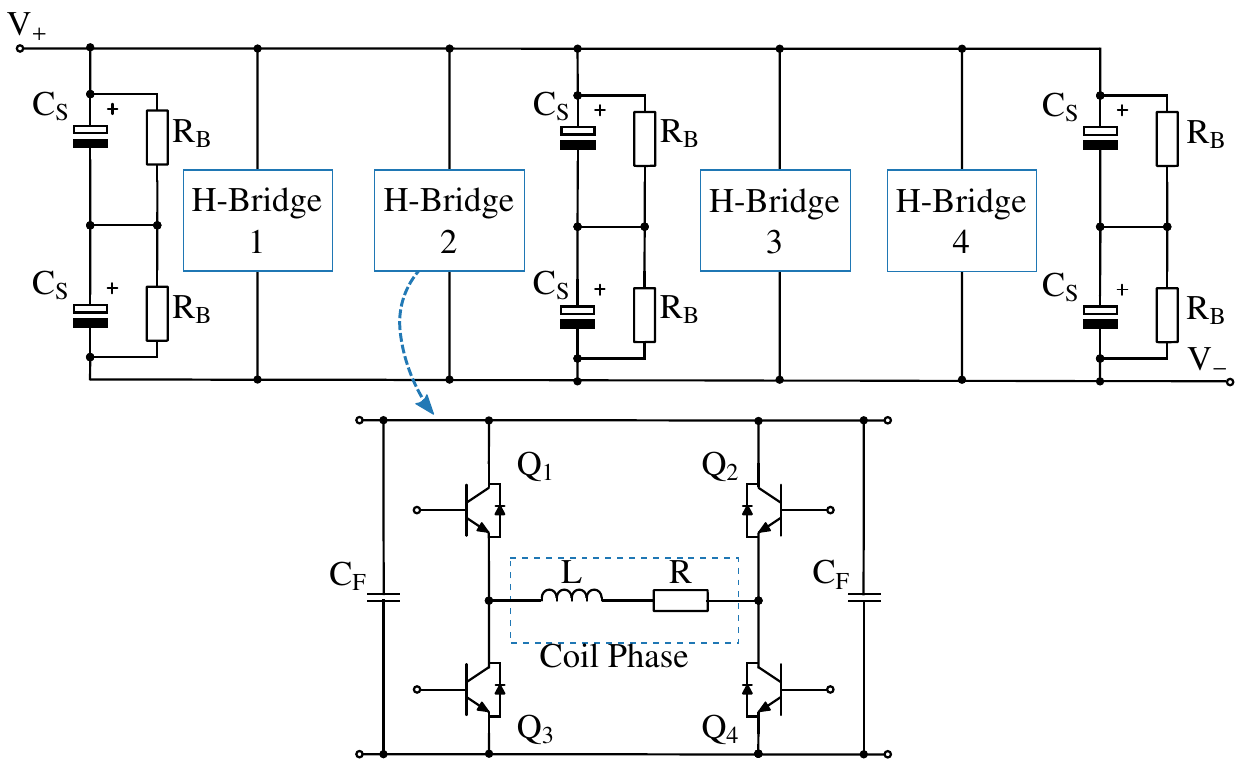}
 \caption{\textbf{The complete circuit schematic of the decelerator electronics.} The circuit diagram shows the power supply and switch circuitry separately. The four H-bridge switch units are constructed from IGBTs and are labelled Q$_\mathrm{i}$ depending on its position within the H-bridge. Furthermore the connections to the coil phase is shown as a resistor, $R$, in series with an inductance, $L$. The filter capacitors, C$_\mathrm{F}$, (1.5 mF rated to 850V) are used to suppress the transient voltage spikes generated while switching the IGBTs. The supply capacitors, C$_\mathrm{S}$, consist of two capacitors (each 2.2mF rated to 450 V) in series to allow the full 800 V to be dropped across them. For completeness, the bleed resistors, R$_\mathrm{B}$ used to discharge the capacitors after use have also been included. Not shown are the driver cards or the connections to the fibre-optic coupler.\label{fig:Dec_circ}}
\end{figure}

The power electronics required for the experiments do not exist commercially so were designed and built in house. The power electronics circuit is shown in Figure~\ref{fig:Dec_circ}. The voltage-switching circuit is an H-bridge, of which there are four, one for each phase of the decelerator coils. The power electronics for the four coil phases are constructed from IGBTs (Semikron SKM400GB12T4) and are labelled Q$_\mathrm{i}$ in Figure~\ref{fig:Dec_circ} depending on its position within the H-bridge. A coil phase is shown as a resistor $R$ and an inductor $L$ in Figure~\ref{fig:Dec_circ}. It should be noted that each of the wires that form a coil phase are wired in series, which ensures the current through the wires in the coil phase are all equal. A capacitor bank consisting of three parallel sets of two 2.2 mF, 450 V capacitors (BHC ALS30A222F420), connected in series and labelled C$_\mathrm{S}$ in Figure~\ref{fig:Dec_circ}), supply the current to the coils. Despite the total capacitance being halved, this allows up to 900 V to be dropped across each pair. The total capacitance is 3.3 mF for each four-H-bridge module. Just as was the case with the quadrupole guide, the capacitors are not fully discharged, which leaves a voltage bias on the capacitor that can be used to drive the current out of the coil faster than it would do if left to free-inductively decay. Also shown in Figure~\ref{fig:Dec_circ} are bleed resistors, $R_\mathrm{B}$, which are there for safety reasons because they ensure that the capacitors are completely discharged after an experimental run. In order to suppress any transient voltage spikes that might damage the IGBTs, 1.5 mF filter capacitors (Vishay MMKP386), labelled C$_\mathrm{F}$ in Figure~\ref{fig:Dec_circ}, are placed in the H-bridge circuitry. Each coil module used in the decelerator, of which there are four, requires its own such power-electronics module. The four power-electronics modules are connected in parallel and charged from a single power supply.

When designing such a PWM power-electronics module it is very important to minimise the stray inductance of the circuit. When a current is passed though the circuit, energy is stored due to this stray inductance. When an IGBT is switched off, a voltage is generated across it to maintain the current in accordance with Faraday's law. In general, the value of stray inductance tends to be on the order of pico-henries, but owing to the rapid change in current the voltage generated may be several hundred volts. This, combined with the supply voltage has the potential to exceed the nominal voltage to which the IGBT device is rated and repeated use under these conditions will eventually destroy the IGBT. Stray inductance can be minimised through careful choice of the circuit layout and includes precautions such as placing the positive and negative supply rails in parallel to each other to reduce the mutual inductance \cite{Mezhiba2002} and careful consideration of the orientation of additional connections.\cite{Grover2002} Additional circuits known as snubber networks can also be used to suppress these voltage spikes.\cite{Mohan2002} The best practice, however, is to optimise the circuit design to keep the stray inductance low.

\begin{figure}
\centering
\includegraphics[width=340pt]{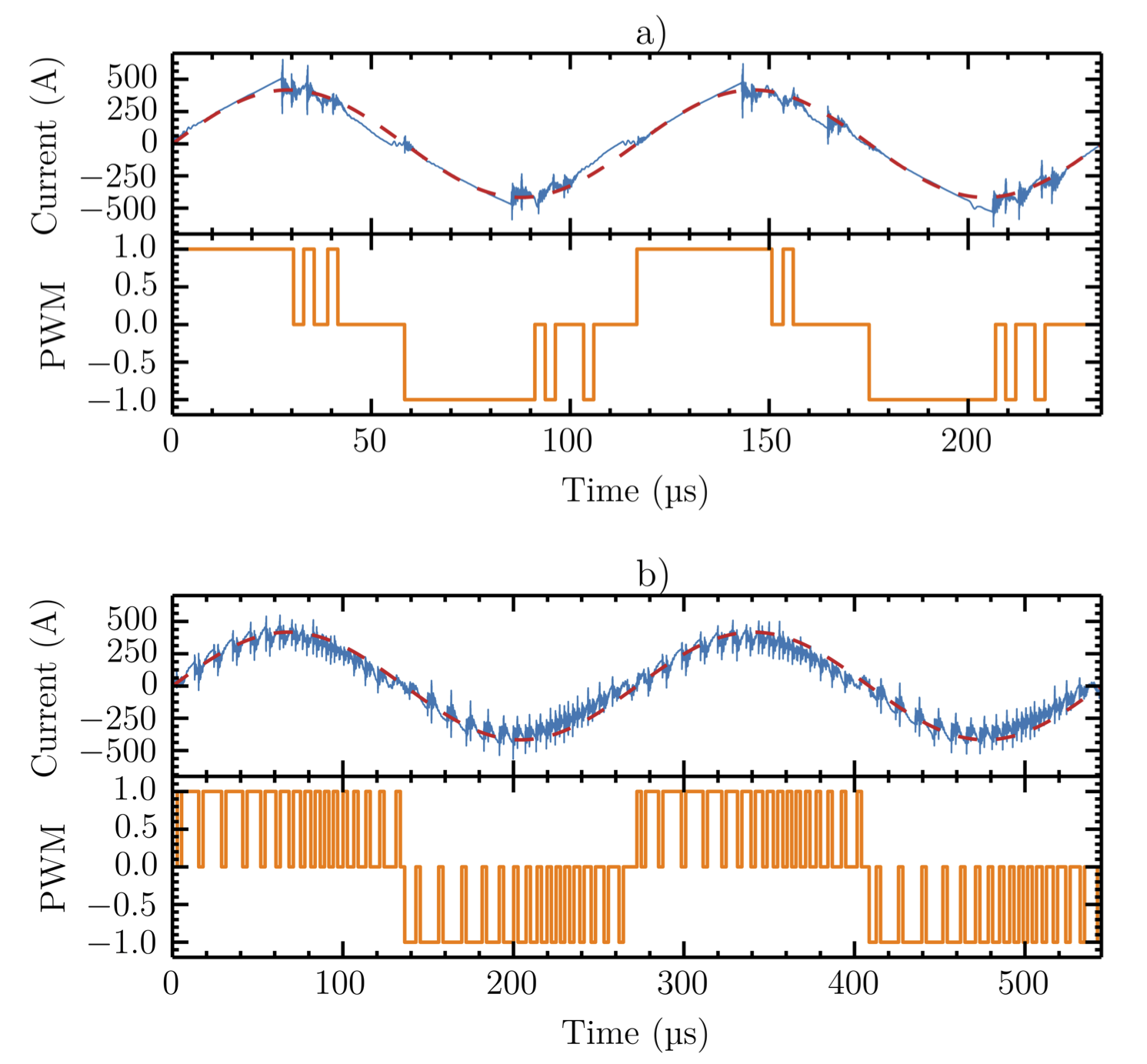}
 \caption{\textbf{Examples of the measured current waveform through one of the coil phases for two different frequencies.} The two examples shown are at frequencies corresponding to a) 350 m$\,$s$^\mathrm{-1}$ and b) 150 m$\,$s$^\mathrm{-1}$. In both cases the reference waveform has been plotted (red dashed). The lower panels of each example gives the PWM signal sent to the driver cards.\label{fig:dec_current}}
\end{figure}

\subsection*{Effects of the PWM} 

The approximate waveforms produced by PWM will effect the properties of the trap as it moves through the decelerator. To investigate this, two current waveforms, which correspond to traps moving at 350~m~s$^{-1}$ and 150~m~s$^{-1}$, shown in Figures~\ref{fig:dec_current}a) and b), respectively, can be used to simulate the time dependence of the field and then compared to corresponding moving traps simulated using pure sinusoidal waveforms. For simplicity, the field produce by the quadrupole guide has not been included. The magnetic field produced by the decelerator coils at an arbitrary point cannot immediately be calculated using the Biot-Savart Law because the wires that form the decelerator are not parallel to any of the coordinate axes in the laboratory frame. The first step in the method to calculate the magnetic field is to rotate the point of interest into the frame of the wire, the angle of the rotation is dependent on the crossing angle of the wire and the $z$ axis and therefore takes values of $\left\lbrace \alpha, \:\pi - \alpha \right\rbrace $, where $\alpha $ = 70$^\circ$ in this case. After this rotation the coordinate system is translated such that the wire now crosses the origin of this new coordinate system so that the Biot-Savart Law for a finite length conductor can now be applied. The field in the laboratory frame at this arbitary point is returned following a second rotation and summing over each wire element that forms a decelerator coil. This method is then repeated for a 3D grid of points, which is then interpolated using the tricubic interpolation method described by Lekien \textit{et al.}\cite{Lekien2005} This allows us to accurately model the field produced by the decelerator.

\begin{figure}[t]
	\centering
	\includegraphics[width=340pt]{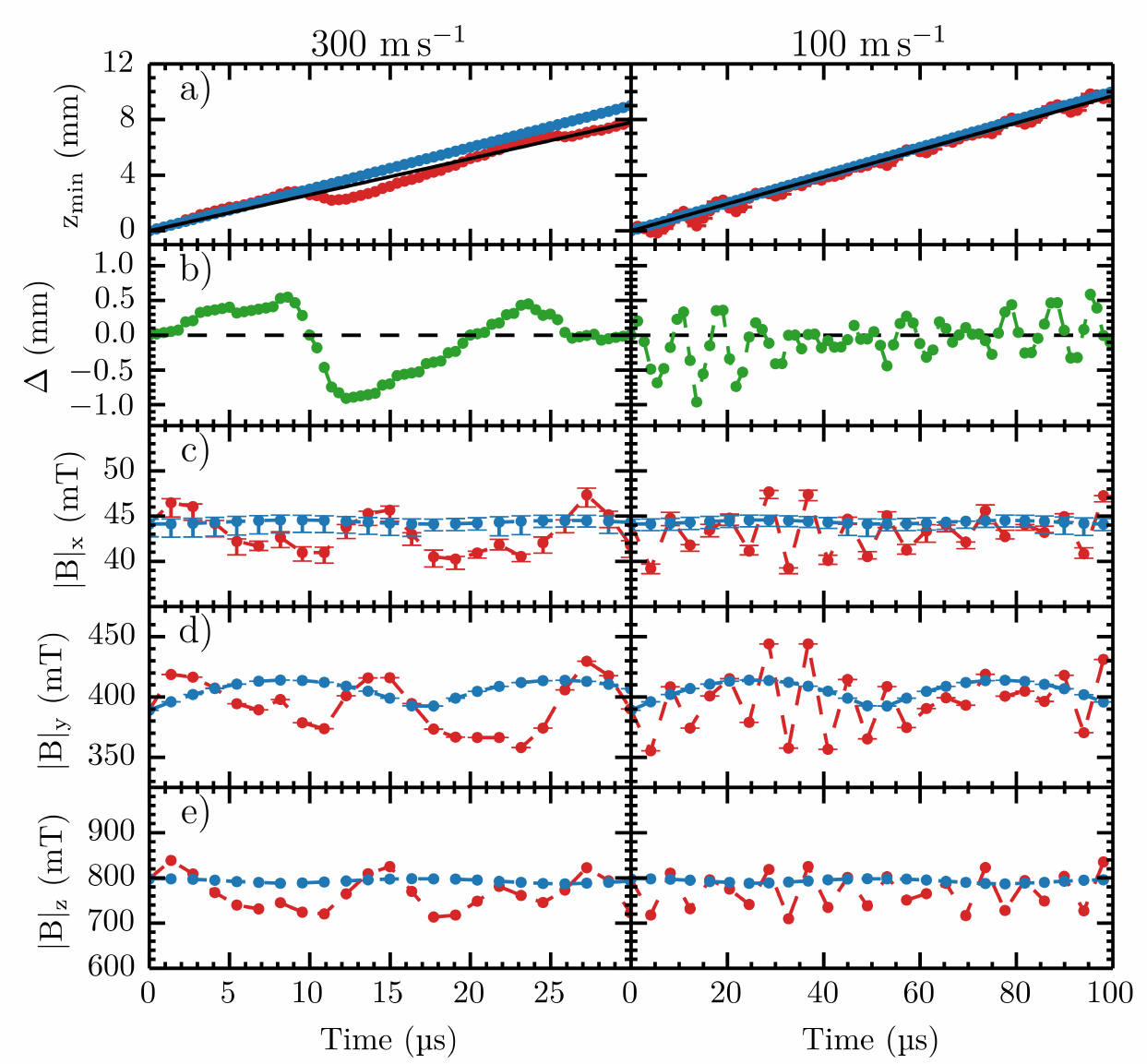}
 \caption{\textbf{The effect of the PWM generated waveform on the properties of the trap for the two frequency waveforms shown in Figure~\ref{fig:PWM}.} Row a) shows the motion of the trap minimum through one quarter period of a coil produced by a synthesised (red) and pure (blue) waveforms for target trap velocities of 300 m$\,$s$^{-1}$ and 100 m$\,$s$^{-1}$. The fitted average trajectory of the trap minimum produced using the synthesised current waveform is also plotted (black). Row b) shows the deviation of the trap minimum produced by the synthesised waveform from the fitted velocity. Rows c), d) and e) shows a comparison of the trap depths as a function of time for the pure (blue) and the synthesised (red) waveforms in the x, y and z directions respectively. The trap depths in the transverse directions are measured at the longitudinal minimum of the trap while the longitudinal trap depth is measured at the front edge of the trap. \label{fig:PWM_analysis}}
\end{figure}

Figure~\ref{fig:PWM_analysis} shows the results of simulations that examined several properties of the trap for the first quarter period of the two current waveforms shown in Figure~\ref{fig:PWM}. Figure~\ref{fig:PWM_analysis}a) shows how the trap minimum varies in position along the $z$ axis with time for the ideal, purely sinusoidal waveform and the PWM-synthesised waveforms. For each velocity, the PWM resulted in a trap velocity lower than the target velocity, 297.4 $\pm$ 0.2 m$\,$s$^{-1}$ and 99.6 $\pm$ 0.4 m$\,$s$^{-1}$, respectively. Figure~\ref{fig:PWM_analysis}b) shows the difference, $\Delta$, in position of the trap minimum along the $z$ axis between the ideal trap trajectory and the actual trajectory. The minimum oscillates around the ideal trajectory with an amplitude of approximately 1 mm. The magnitude of the oscillations is slightly smaller for waveforms that contain more switching events and the frequency of the oscillations is higher.

The trap depths in the $x$, $y$ and $z$ directions are plotted for the ideal sinusoidal waveform and for the PWM waveform in Figures~\ref{fig:PWM_analysis}c), d) and e), respectively. The depth of the trap in the $x$ and $y$ directions have been calculated at the position of the trap minimum in the $z$ direction. In the $z$ direction, the heights of the front and rear edges of the trap vary independently, although they do average to the same height over time. The trap depths at a specific point in time are in reference to the front trap edge rather than the rear trap edge because the packet of atoms resides on the front edge due to deceleration. The synthesised waveforms introduce fluctuations in the trap depth of approximately 10 mT, 80 mT and 130 mT in the $x$, $y$ and $z$ directions, respectively, for both frequencies considered here. The latter is approximately 5 $\%$ of the average field depth. The pure sinusoidal waveforms also exhibited periodic oscillations in the trap depth, with frequencies of 58.8 kHz for the 300 m$\,$s$^{-1}$ waveform and 19.6 kHz for the 100 m$\,$s$^{-1}$ waveform and with amplitudes of 0.5 mT, 25.0 mT and 11.4 mT in $x$, $y$ and $z$, respectively, for both velocities studied. This is due to the small number of overlapped traps in a period of the coil (see Figure~\ref{fig:coil_geometry}).

In general, the quality of the synthesised waveform directly influences the dynamics of the trap. A lower quality waveform (i.e. one with a few current pulses) tends to have a slower mean velocity than that of the ideal waveform and exhibits larger oscillations about the ideal trajectory compared to a higher quality waveform. A higher quality waveform will experience the same variation of the field depths as a lower quality waveform, but because the switching events occur more frequently, the average field depth tends towards that of the pure synthesised waveform. As a result the decelerator benefits from lower trap velocities. In the next section, the effect of PWM on the phase-space acceptance of the moving trap will be investigated using Monte-Carlo trajectory simulations.

\section*{Phase space acceptance}

To understand how the phase-space acceptance of the decelerator varies under a range of deceleration conditions, 3D Monte Carlo trajectory simulations have been run using a time-averaged magnetic field and an ideal, sinusoidal waveform to calculate the phase-space acceptance. The justification for using a time-averaged magnetic field is that the characteristic frequency of, for example, a metastable argon atom oscillating in the trap does not exceed a few kHz. An analysis of the trap variations caused by the PWM reveals that the trap varies within a frequency range of 40 to 180 kHz depending on the number of switching events present in the current waveform. This difference in frequencies might be expected to be large enough that the atoms cannot react fast enough to the rapidly changing field but instead feel the average potential.\cite{Landau1969} The effect of the PWM on the phase-space acceptance is discussed later.

\begin{figure}[t]
	\centering
	\includegraphics[width=340pt]{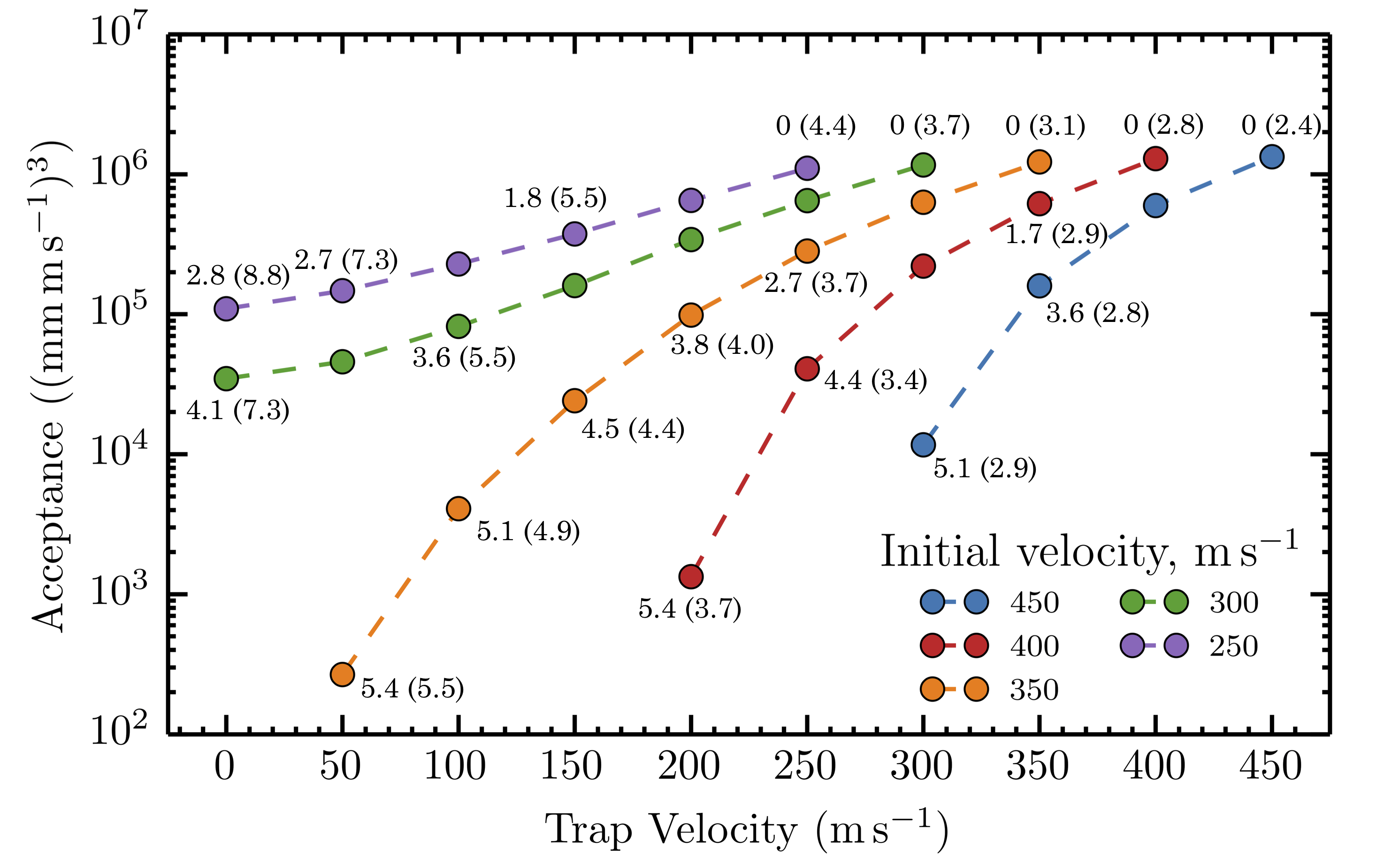}
	\caption{\textbf{The 6D phase-space acceptance of the decelerator as a function of the final trap velocity for a series of initial trap velocities.} The acceptance of the decelerator operating at 1000 A peak with a quadrupole field generated with 700 A. Here an ideal nine-module long, 1.10~m, decelerator has been used to illustrate the dependence of the acceptance of the decelerator on the initial trap velocity for a range of final trap velocities. The packets of metastable argon atoms have an initial velocity in the range of 250 to 450 m$\,$s$^\mathrm{-1}$. Each data point is labelled with the value of the deceleration, in units of $10^4$~m~s$^{-2}$, and (in parentheses) the time the metastable argon atoms spend in the decelerator in units of ms.}\label{fig:Dec_initAcc}
\end{figure}

The simulations began with 1,000,000 metastable argon atoms uniformly distributed over a spatial volume of 5.3 mm $\times$ 5.3 mm $\times$ 22 mm and a velocity-space volume of 20 m$\,$s$^\mathrm{-1}$ $\times$ 40 m$\,$s$^\mathrm{-1}$ $\times$ 60 m$\,$s$^\mathrm{-1}$. This gives a 6D-phase-space volume of $3.0\times10^7$~mm$^3$\,(m~s$^{-1}$)$^3$. During the simulation, atoms were deemed to be lost once they passed beyond a trap maximum or when they reached the internal wall of the vacuum tube. Several deceleration scenarios were considered with starting velocities in the range 250 to 450~m~s$^{-1}$ and with finishing velocities in the range 0 to 300~m~s$^{-1}$ over a decelerator length of 1.10~m. A peak current of 1000~A was assumed. The initial velocity of the moving trap was set to be equal to the initial mean velocity of the gas packet in the $z$ direction. The phase-space acceptance was calculated by multiplying the initial 6D-phase-space volume by the proportion of the initial number of atoms remaining in the trap at the end of the decelerator. Figure~\ref{fig:Dec_initAcc} shows the results of the simulations.

For simulations where no deceleration was applied (3D guiding), the phase-space acceptance was approximately $10^6$~mm$^3$\,(m~s$^{-1}$)$^3$ dropping slightly as the starting velocity reduced. This drop was due to the atoms spending longer in the decelerator giving atoms more time to explore phase-space and to find a way out of the trap. This trend is seen throughout the data. For example, the phase-space acceptance for deceleration from 350~m$\,$s$^\mathrm{-1}$ to 250~m$\,$s$^\mathrm{-1}$ is higher than the the phase-space acceptance for deceleration from 250~m$\,$s$^\mathrm{-1}$ to 50~m$\,$s$^\mathrm{-1}$, despite the deceleration being the same, because in the latter case the atoms spend an extra 2.1 ms in the decelerator. Clearly, the phase-space acceptance is larger when the deceleration is lower and when the deceleration takes less time. In order for the drop in phase-space acceptance between the cases of 3D guiding and deceleration to remain lower than one order of magnitude, the deceleration should be kept below approximately $3\times10^4$ to $4\times10^4$~m$\,$s$^\mathrm{-2}$.

\begin{figure}
\centering
	\includegraphics[width=360pt]{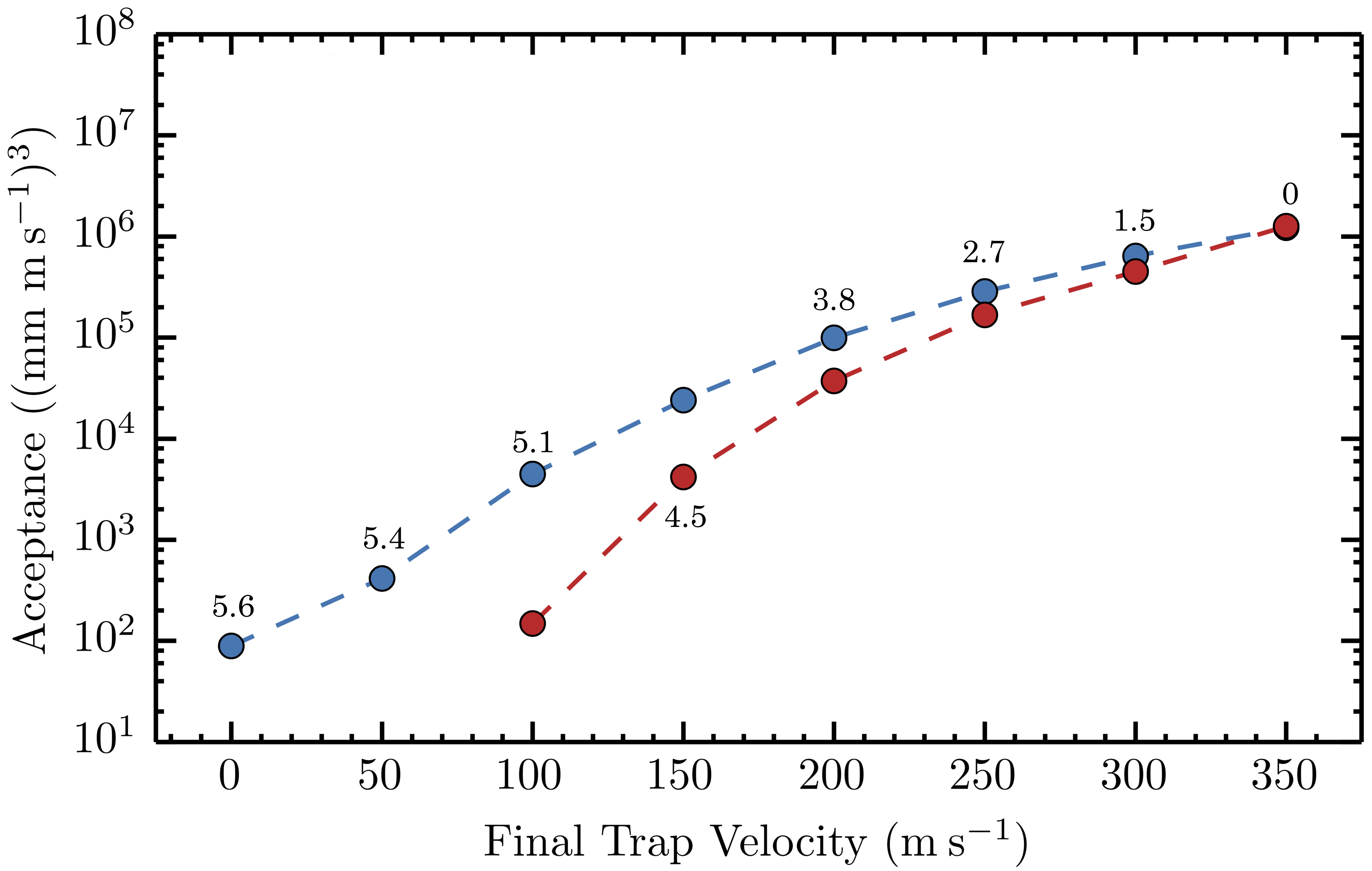}
    \caption{\textbf{The acceptance of the decelerator as a function of final trap velocity in the deceleration mode.} The acceptance of the decelerator as a function of final trap velocity for a nine-module long decelerator operating at 1000 A peak. The plot include the acceptance for a decelerator using the pure current waveform (blue) and synthesised current waveform (red). The mean longitudinal velocity of the gas packet was 350 m$\,$s$^\mathrm{-1}$. Initially, the trap velocity was set to 350 m$\,$s$^\mathrm{-1}$, thus the magnitude of the deceleration ranges from 0 to 5.56 $\times$ 10$^4$ m$\,$s$^\mathrm{-2}$. Each data point is labelled with the value of the deceleration, in units of $10^4$~m~s$^{-2}$.}\label{fig:Deceleration_Acceptance}
\end{figure}

Simulations equivalent to those described above that accounted for oscillations in the trap shape and trajectory due to PWM would require an unreasonable amount of time to compute because the magnetic field would have to be calculated and interpolated for every time step in the simulation instead of only once at the beginning of the simulation. Several simplifications to the simulation process have been made to capture the effect the PWM has on the phase-space acceptance. Firstly, the trap depth and trap position were calculated along the $z$ axis only for every time step in the simulations. This is an approximation because the field along the $z$ axis is influenced, at the level of a few percent, by the fields in the $x$ and $y$ directions. Deceleration from 350~m$\,$s$^\mathrm{-1}$ to a range of final velocities from 350~m$\,$s$^\mathrm{-1}$ to zero was simulated over a decelerator length of 1.10~m and with a peak current of 1000~A. For each scenario, the time-dependent deviations of the trap depth and position from the average trap depth and position were approximated with sine functions. A trap with the average trap depth was then calculated in 3D and its trap depth was varied using a scaling factor according to the sine function. 3D Monte Carlo trajectory simulations were then run for each scenario using traps that oscillated in position and in depth with amplitudes and frequencies set by the sine functions. Figure~\ref{fig:Deceleration_Acceptance} shows the resulting phase-space acceptances compared to the results when the effects of PWM were not taken into account. For decelerations of approximately $3\times10^4$~m$\,$s$^\mathrm{-2}$, PWM causes the phase-space acceptance to drop by approximately 40\%. This drop gets worse as the deceleration is increased and is a much as 95\% when the deceleration is approximately $5\times10^4$~m$\,$s$^\mathrm{-2}$. The explanation for this trend is simple. For higher decelerations, the packet of atoms resides higher up the leading edge of the trap. This is the region of the trap where oscillations in the magnetic field caused by the PWM are greatest.
Overall, deceleration should be kept below approximately $3\times10^4$~m$\,$s$^\mathrm{-2}$.

\section*{Experimental results}

\begin{figure}
\centering
	\includegraphics[width=360pt]{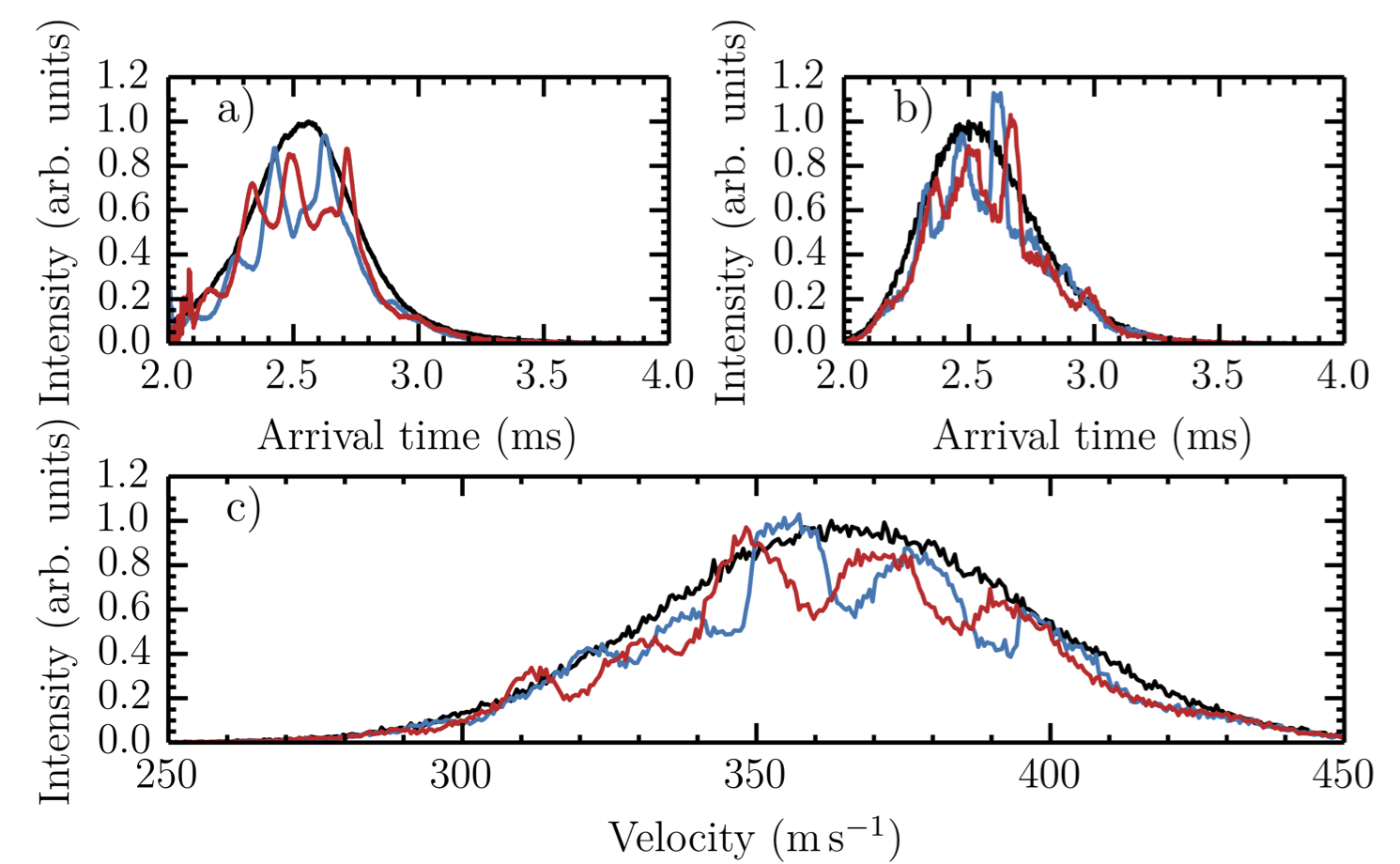}
    \caption{\textbf{Comparison of the experimental and simulated TOF profiles while the decelerator is in the 3D guiding mode. } Panels (a) and (b) show the TOF profiles for the experimental and simulated data respectively for average trap velocities of 351 (red) and 373 (blue) m$\,$s$^\mathrm{-1}$. The zero field TOF profile has been plotted for reference in each instance (black). Panel (c) shows the velocity distribution taken from the simulated data as the particles reach the MCP.}\label{fig:3D_guiding_velocity}
\end{figure}

Figure~\ref{fig:3D_guiding_velocity} shows experimental results with the decelerator in 3D guiding mode. In this mode of operation, the 3D trap is not decelerated so the velocity is constant at a set value. Figure~\ref{fig:3D_guiding_velocity}a) shows the measured arrival time distribution for Ar($^3P_2$) atoms arriving at the MCP for trap velocities of 351 m~s$^{-1}$ (red data) and 373 m~s$^{-1}$ (blue data). The peaks in arrival time correspond to atoms that have been prevented from spreading out by the moving traps. Figure~\ref{fig:3D_guiding_velocity}b) shows the results of simulations of the 3D guiding experiments for both velocities. The agreement between experiment and simulations is good. Figure~\ref{fig:3D_guiding_velocity}c) shows the velocity distributions obtained from the simulations of both experiments. There is a peak centred at approximately 349~m~s$^{-1}$ in the red data and a peak centred at approximately 375~m~s$^{-1}$ in the blue data. The widths of the peaks are consistent with the predicted range of velocities that can be trapped.

The peak arrival times of trapped atoms deviate slightly in the simulations compared to the corresponding experimental peak arrival times. The greatest contributing factor to this error is due to the approximate way in which the PWM affects the motion of the trap is treated in the simulations. Peaks in the simulations tend to be more intense than experimental peaks. It is likely that this is due to the fact that the potentials are switched on instantaneously in the simulations whereas in the experiments the potential ramp up in a finite time causing atoms to be lost from the trap.

In these experiments, the decelerator was not performing optimally. A safety feature of the decelerator is that the charge in the capacitors in the coil-driver power-electronics modules must be discharged quickly to ground through high-current contactors in the event of an emergency. The contactors for each of the four coil sections were faulty. Consequently, as the first module discharged, the other modules charged the partially discharged first module through its faulty contactor, This caused the peak current that could be drawn by the next coil section to be reduced because the capacitors in its driver module were never fully charged. This repeated to varying degrees down the decelerator meaning that although each coil section was supposed to draw a peak current of 500~A, in reality the coils could only draw peak currents in the range of 100--200 A. This effect was taken into account in the simulations. However, at the boundary between two coil sections operating at different peak currents, the simulations assumed an instantaneous transition from one section to the other; in reality the transition was not so seamless. This contributed to the error between experiments and simulations.

\begin{figure}
\centering
	\includegraphics[width=360pt]{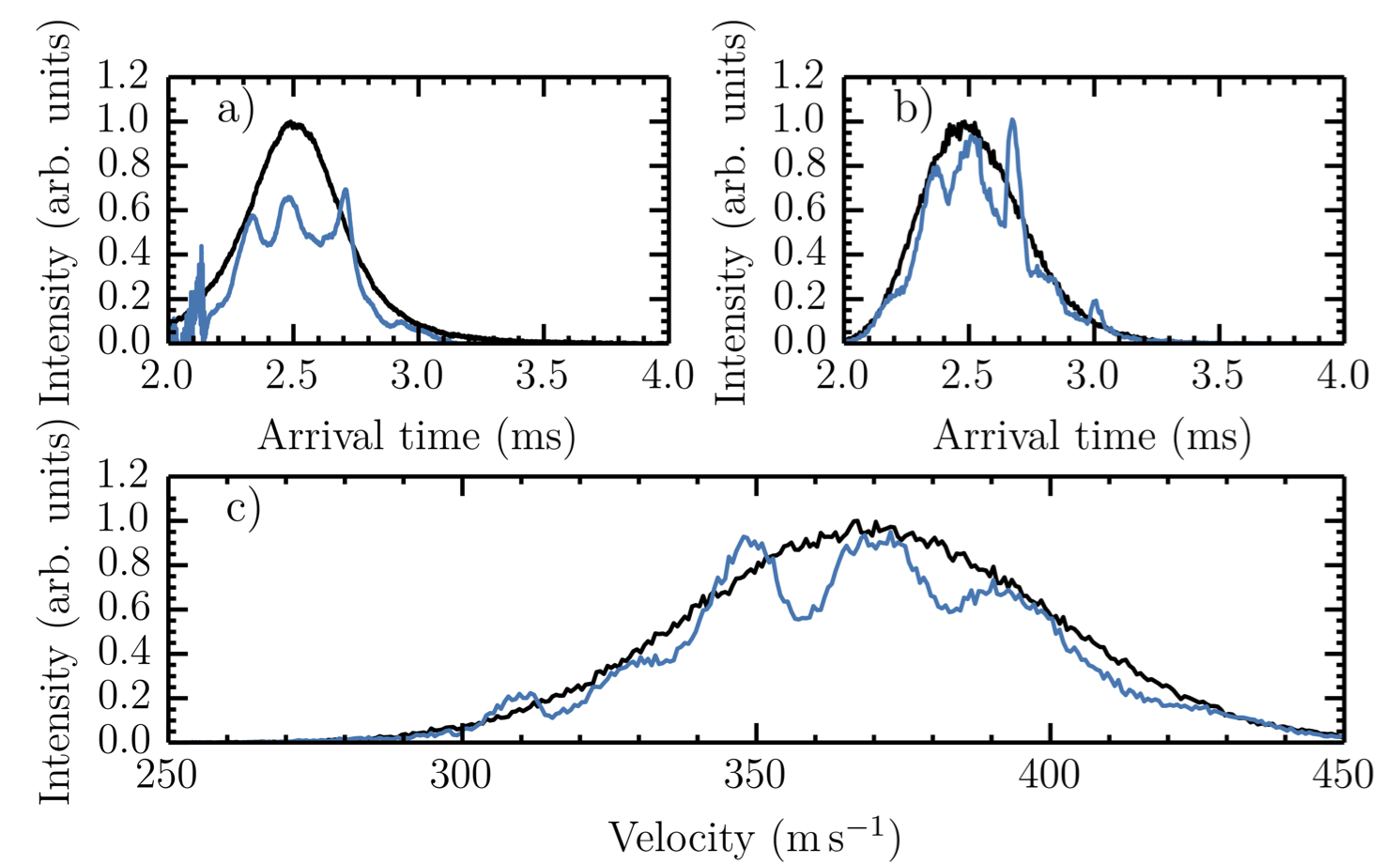}
    \caption{\textbf{Comparison of the experimental and simulated TOF profiles while the decelerator is in the deceleration mode.} Panels (a) and (b) show the TOF profiles for the experimental and simulated data respectively for a trap decelerated from 343~m~s$^{-1}$ 304~m~s$^{-1}$ (blue). The zero field TOF profiles has been plotted for reference (black). Panel (c) shows the simulated velocity distribution.}\label{fig:deceleration}
\end{figure}

Figure~\ref{fig:deceleration}a) shows the experimental results with the decelerator operating in deceleration mode where the initial trap velocity was 342~m~s$^{-1}$ and the final trap velocity was 304~m~s$^{-1}$. Therefore, the deceleration has removed 21\% of the kinetic energy of the trapped atoms in the beam. The peak in the deceleration data with a final trap velocity of 304 m$\,$s$^\mathrm{-1}$ has an arrival time of 2.71 ms whereas the corresponding peak in the simulated data arrives at 2.68 ms, which is why the peak in the simulated velocity distribution appears centred at 310~m~s$^{-1}$. This discrepancy between experiment and simulation is due to the effects described above. In an optimally performing decelerator working at a peak current of 500~A the deceleration will clearly be more efficient.

\section*{Conclusions and outlook}
A type of moving-trap Zeeman decelerator (MTZD) for use in molecule trapping experiments has been presented. MTZDs are designed to maximise the transmitted molecule numbers by mitigating the losses due to overfocusing encountered in multistage Zeeman decelerators. Such losses drastically reduce the 6D phase-space acceptance at velocities less than \emph{ca.}~100~m~s$^{-1}$. Mitigation is done by creating a moving, 3D magnetic trap, as opposed to the time-averaged moving trap in multistage Zeeman decelerators, for the entire duration of the deceleration process all the way down to a standstill, which is ideal for molecule trapping. The MTZD is based on a design published in Trimeche \emph{et al.}.\cite{Trimeche2011} The Trimeche decelerator has a large 3D velocity acceptance, but a very small 3D velocity acceptance. The design of the MTZD presented here has been significantly modified to greatly increase the 3D spatial acceptance without cost to the 3D velocity acceptance thus ensuring a large 6D phase-space acceptance.

We have presented the technical design and operation principles of the MTZD and studied its properties with extensive Monte-Carlo trajectory simulations. We have detailed the principles of the custom power electronics that are required to drive switchable DC currents up to 700~A through the 2D quadrupole guide, which provides the transverse part of the moving 3D magnetic trap. We also detailed the custom power electronics that are required to drive an alternating current of up to 1000~A peak through the deceleration coils, which provide the longitudinal part of the moving 3D magnetic trap. We have described the pulse-width modulation (PWM) technique used in the AC power electronics to generate high-frequency sinusoidal waveforms and discussed the effect that deviations from a perfect sinusoid have on the movement of the 3D magnetic trap. With simulations we have shown how the 6D phase-space acceptance of the trap varies in a 1.1~m decelerator operating at maximum currents in the 2D quadrupole guide and in the deceleration coils under different deceleration scenarios. With PWM taken into account, we concluded that decelerations  should be kept below $3\times10^4$~m~s$^{-2}$ to maintain a good 6D phase-space acceptance.

In a proof-of-principle experiment, we demonstrated the implementation of the MTZD design by 3D guiding at constant velocity and by decelerating beams of metastable argon atoms in a relatively short, 0.49~m decelerator. The decelerator ran with varying peak currents in the range of 100--200~A due to faulty components, but we achieved 3D guiding at 373~m~s$^{-1}$ and deceleration from 342~m~s$^{-1}$ to 304~m~s$^{-1}$, which corresponds to the removal of 21\% of the kinetic energy of the beam. The experimental TOF profiles show good agreement with simulations.

The immediate outlook is to replace the faulty components to return the decelerator to full operating specifications. With the current decelerator, we can, in principle, bring beams of paramagnetic atoms or molecules to a stanstill, providing the mass-to-magnetic-moment ratio is below around 8~u~$\mu_\mathrm{B}^{-1}$, which includes H atoms, O atoms and NH radicals. Theoretical calculations suggest that ultracold H atoms are a versatile sympathetic coolant for molecules.\cite{GonzalezMartinez2013} H atoms are very difficult to cool into the ultracold regime, but if the sympathetic cooling prediction is turned on its head then H atoms should be sympathetically coolable with laser coolable alkali metals such as Li, which would open the door to producing a H-atom fountain for precision measurements. Akerman \emph{et al.} have already shown that an MTZD can be used to simultaneously load a magnetic quadrupole trap with O$_2$ molecules and Li atoms, which suggests that similar can be done with H and Li atoms. However, magnetic quadrupole traps are not suitable for laser cooling due to the large magnetic field gradients required to trap the mK temperature atoms and molecules that the MTZD produces. A new type of magnetic trap in which molecules can be trapped \emph{and} in which the co-trapped atoms can be laser cooled is required.

We are planning to extend the decelerator to 1.10~m by adding an additional five deceleration-coil sections and to combine the decelerator with a buffer-gas beam source capable of producing intense, short-pulse beams at velocities as low as 125~m~s$^{-1}$.\cite{Truppe2017(1)} This would open the door to the deceleration of molecules with mass-to-magnetic-moment ratios as high as 100~u~$\mu_\mathrm{B}^{-1}$ for trapping and sympathetic cooling experiments.



\section*{Acknowledgements}

For their technical assistance, we are grateful to the Mechanical and Electronics Workshops, especially John F. Scott, in the Physics and Chemistry Departments at Durham University and to the Mechanical Workshop at Imperial College. We also thank Barry Moss for his assistance with the design of the power electronics. This work was supported by the EPSRC under the Programme Grant EP/I012044/1.

\end{document}